\documentclass[twocolumn,aps,prl,groupedaddress]{revtex4-2}
\usepackage{mystyle}

\newcommand{\synGAIN}[3]{\mbox{\Large$\Sigma$}_{#1}[{#2},{#3}]} % 合成反応の収量
\newcommand{\decGAIN}[3]{\mbox{\Large$\Sigma$}'_{#1}[{#2},{#3}]} % 分解反応の収量
\newcommand{\syngain}[3]{\Sigma_{#1}\mbox{\footnotesize$[{#2},{#3}]$}} % 小さいバージョン
\newcommand{\decgain}[3]{\Sigma'_{#1}\mbox{\footnotesize$[{#2},{#3}]$}}
\newcommand{\SECTION}[1]{\paragraph*{#1.\rule[3pt]{10pt}{0.2pt}}} % 斜体+横線で章を分ける形式
\newcommand{\EXAMPLE}{XABA} % 例として考えている分子鎖
\newcommand{\PROTOCOL}{[c_A(t), c_B(t)]_{0\leq t\leq T}} % 操作プロトコル
\newcommand{\PAPER}{Letter} % この論文の呼称

\begin{document}

% タイトルと著者名・所属
\title{Characterizing the asymmetry in hardness between synthesis and destruction of heteropolymers}
\author{Ikumi Kobayashi}
\author{Shin-ichi Sasa}
\affiliation{Department of Physics$,$ Kyoto University$,$ Kyoto 606-8502$,$ Japan}

\date{\today}

% 概要
\begin{abstract}
We present a simple model describing the assembly and disassembly of heteropolymers consisting of two types of monomers $A$ and $B$.
We prove that no matter how we manipulate the concentrations of $A$ and $B$, it takes longer than the exponential function of $d$ to synthesize a fixed amount of the desired heteropolymer, where $d$ is the number of $A$-$B$ connections.
We also prove the decomposition time is linear for chain length $n$.
When $d$ is proportional to $n$, synthesis and destruction have an exponential asymmetry.
Our findings may facilitate research on the more general asymmetry of operational hardness.
\end{abstract}
\maketitle

% はじめに -------------------------------------------------------
\SECTION{Introduction}
% 第1段落：自然現象における実現が簡単な過程と難しい過程
It is hard to construct a complex system, while it is easy to destroy it \cite{Deutsch_2013}. A typical example of such a system is a living organism \cite{Marletto_2015}. Even though various kinds of damage may stop biological activity, it is almost impossible to bring an organism back to life once it has died \cite{Bernat_2013}. One possible cause of the irreversibility of biological death is the asymmetry in the difficulty of assembling and destroying structures. To build a structure with some desirable function or property, each component must be carefully arranged in a specific order but no such care is required to destroy it.  In this {\PAPER}, we attempt to characterize such asymmetry.

% 第2段落：先行研究
A similar asymmetry is formulated in computational complexity theory, where problems are classified by the amount of computational time and memory required to solve a given problem using a model computer, such as Turing machine \cite{Arora_2009}. Concretely, the concept of NP and P characterizes the asymmetry of computation. If a problem is in NP and not P, we can efficiently (i.e., in polynomial time of the length of the input bit sequence) determine whether a given candidate is a suitable solution to the problem or not, but it takes an exponentially long time in the worst case to find the solution itself  \cite{Moore_2011}. 
To the present, there have been many attempts to link the concepts of the theory of computation to the description of natural phenomena \cite{Barahona_1982, Bennett_1982, Mezard_2001, Monasson_1999, Unger_1993, Zurek_1989, Fu_1986, Cubitt_2015, Hjelmfelt_1992, Magnasco_1997, Abrams_1998, Lucas_2014, Kari_2008} and this would be an interesting research direction.

% 第3段落：組み立てと破壊の非対称性を理解するための最小模型
Motivated by these studies, in this {\PAPER}, we formulate the operational asymmetry between synthesis and destruction of one-dimensional heteropolymers. 
Because we restrict our analysis to a class of simple systems with an experimentally accessible setting, there is no direct correspondence to computational complexity theory.
Nevertheless, we can show that there exists a fundamental limit to the synthesis rate that cannot be exceeded by any external operations in that setting. 
The results obtained for the specific model may provide useful insights into more general studies on the cost of polymer synthesis or protein synthesis under nonequilibrium environments.

% 第4段落：今回構成した模型の説明と帰結
We introduce an idealized description of the synthesis and decomposition of a one-dimensional molecular chain comprising two types of molecular subunits. 
The concentrations of molecular subunits are controlled by an external operator.
The difficulty of synthesis (destruction) is quantified by the time $T$ it takes to synthesize (destruct) a fixed amount of molecular chains with the desired sequence.
In the synthesis process, we prove that $T$ is greater than an exponential function of $d$ no matter what external operations are performed, where $d$ is the number of connections between different types of molecular subunits.
The difficulty of synthesizing heteropolymers in this system is characterized by this quantity $d$.
We also prove that the time required to decompose a fixed amount of molecular chains is smaller than a linear function of the chain length $n$.
For molecules where $d$ is proportional to $n$, there is an exponential asymmetry between the hardness of synthesis and destruction.

% 合成反応の設定
\SECTION{Synthesis process}
We consider the synthesis of a one-dimensional molecular chain by the sequential binding of molecules $A$ and $B$ to reaction nucleus $X$ as displayed in Fig. \ref{fig:Synthesis_Destruction}.  An external operator aims to synthesize the desired molecular chain as efficiently as possible by controlling the concentrations of $A$ and $B$ over time.

We assume the only possible reaction is the binding of molecule $Z\in \SET{A, B}$ to a molecular chain $M$ to form $MZ$, 
\EQ{ \label{gousei_reaction}  M+Z\xrightarrow{k_Z}MZ,}
where the rate constant $k_Z$ depends only on $Z$ \cite{Note1}. For simplicity, we ignore the reverse reaction in which a bound molecule is detached. The reaction tank is connected to particle reservoirs that separately supply $X$, $A$, and $B$ (see Fig. \ref{fig:Setup}(a) for an illustration). The concentration of each chemical species and the temperature in the reaction chamber are spatially homogeneous, and the numbers of each molecule are  large enough that the behavior of the system is well described by the following deterministic rate equations:
\MULT{ \label{gousei_rate_eq_M}
\DIFF{t}c_{MZ}(t) = -k_Ac_A(t)c_{MZ}(t)-k_Bc_B(t)c_{MZ}(t)\\+k_Zc_Z(t)c_M(t),}
\vspace{-0.7cm}
\MULT{ \label{gousei_rate_eq_X} 
\DIFF{t}c_X(t) = -k_Ac_A(t)c_{X}(t)-k_Bc_B(t)c_{X}(t) + J_X,}
where $c_M(t)$ is the concentration of molecule $M$ at time $t$.
The first and second terms on the right-hand side of Eq. (\ref{gousei_rate_eq_M}) correspond to the reactions $MZ + A \to MZA$ and $MZ + B \to MZB$, respectively; the third term represents the contribution from the reaction $M + Z \to MZ$. 

\footnotetext[1]{Physically, this assumption corresponds to a situation in which the structure of the binding site at the end of the molecular chain does not change, regardless of the type of the previously bound molecule.}

The concentration dynamics in the reaction chamber can be divided into three types. First, the concentrations of molecules other than $X, A, B$ evolve in time according to Eq. (\ref{gousei_rate_eq_M}). Second, the dynamics of $c_X(t)$ are described by Eq. (\ref{gousei_rate_eq_X}), where the last term represents the constant supply of reaction nucleus $X$ from the particle reservoir to the reaction tank. Third, we assume that the concentrations of molecules $A$ and $B$ are controllable in time by an external operator.

\FIG{Synthesis_Destruction}{fig:Synthesis_Destruction}
{Schematic illustration of the synthesis and destruction of a one-dimensional molecular chain $\EXAMPLE$. The green triangle, blue circle, and red square represent the reaction nucleus $X$, molecule $A$, and molecule $B$, respectively. We assume that the only possible reactions are the sequential binding or detachment of the molecules. }

\FIG{Setup}{fig:Setup}
{Schematic illustration of the setup.
(a) Setup for synthesis process. The reaction tank is connected to particle reservoirs that supply reaction nucleus $X$ and material molecules $A$ and $B$. Molecule $X$ is supplied at a constant rate $J_X$. The operator's task is to synthesize the desired molecule ($\EXAMPLE$ in this example) as efficiently as possible by controlling the concentrations $c_A(t)$ and $c_B(t)$.
(b) Setup for destruction process. The concentration of the molecular chain to be decomposed ($\EXAMPLE$ in this example) is kept constant and $X$ is recovered to the particle bath at rate $\tau_X^{-1}c_X(t)$. The total amount of $X$ recovered is $\decgain{X}{0}{T}$. 
}

At time $t = 0$, the reaction chamber is empty. The total amount of molecule $MZ$ synthesized by reaction (\ref{gousei_reaction}) from $t = 0$ to $t = T$ is
\EQ{ \label{Sigma_def}
\synGAIN{MZ}{0}{T} \equiv \INT{0}{T}{dt}{k_Zc_Z(t)c_M(t)}. }
Let $d$ be the number of connections between different types of molecular subunits. For example, $d = 2$ for $XABBA$ and $d = 5$ for $XAABABBAB$.

% 合成反応の結果
The first main result of this {\PAPER} is that the synthesis of a constant amount of the molecular chain takes an exponentially long time to the molecular length $n$ unless $d = \order{\log n}$, no matter how the concentrations of material molecules are manipulated. Specifically, we can prove that the time $T$ required to synthesize a certain amount $\syngain{M}{0}{T}$ of molecular chain $M$ satisfies
\EQ{ \label{T_gousei} T \;\;>\; \frac{\syngain{M}{0}{T}}{J_X}\times 2^{d/3}}
for any synthesis protocol $\PROTOCOL$.

% 分解反応の設定
\SECTION{Destruction process}
Here, we consider processes that fragment the molecular chain into monomers $X$, $A$, and $B$ by sequential detachment of $Z\in \SET{A,B}$ as displayed in Fig. \ref{fig:Synthesis_Destruction}. The possible reactions are desorption of material molecule $Z$ from molecular chain $MZ$, 
\EQ{ \label{bunkai_reaction} MZ\xrightarrow{k'_Z}M+Z.}
Similar to  synthesis process, we assume that the rate constant $k'_Z$ depends only on $Z$ and that the reverse reaction (recombination) does not occur. The rate equations are then 
\EQ{ \label{bunkai_rate_eq_M}
\DIFF{t} c_{MZ}(t) = -k'_{Z}c_{MZ}(t) + k'_{A}c_{MZA}(t) + k'_{B}c_{MZB}(t),}
\EQ{ \label{bunkai_rate_eq_X}
\DIFF{t} c_X(t) = -\tau_X^{-1}c_X(t) + k'_{A}c_{XA}(t) + k'_{B}c_{XB}(t),}
where $\tau_X^{-1}$ is the recovery rate constant of $X$ from the reaction vessel to the particle bath (see Fig. \ref{fig:Setup}(b) for an illustration). The concentration of the molecular chain to be decomposed $c_M$ is assumed to be constant.

At $t = 0$, the reaction chamber is empty. The total amount of molecular chains completely disassembled from $t = 0$ to $t = T$ is equal to the total amount of reactant nuclei $X$ recovered in the particle reservoir: 
\EQ{ \decGAIN{X}{0}{T} \equiv \INT{0}{T}{dt}{\tau_X^{-1} c_{X}(t)}.}

% 分解反応の結果
We can then show that a linear time with respect to the molecular length $n$ is sufficient to decompose a constant amount $\decgain{X}{0}{T}$ of the molecular chain. That is,
\EQ{ \label{T_bunkai} T\, < \,\,2\overline{\tau}\times n + \FRAC{2\tau_X\decgain{X}{0}{T}}{c_M},}
where $\overline{\tau}$ is the average molecular detachment time.

% 導出の流れ　-----------------------------------------------
\SECTION{Outline of the derivation}
% 合成に指数時間を要することの証明の流れ
First, we explain the outline of the proof that the synthesis reaction takes an exponentially long time to the length of the molecular chain. We focus on the processes where different types of molecules are connected. Given a specific synthesis protocol $\PROTOCOL$, we can express $c_{MZAB}(t)$ as a function of $c_M(t)$ for any $Z\in \SET{A,B}$ and show that $\synGAIN{MZABZ^*}{0}{T} < \synGAIN{MZ}{0}{T} \times 1/2$ for any $Z, Z^* \in \SET{A,B}$.
Because this relation applies to each connection \cite{Note3} between $A$ and $B$ in the molecular chain, the upper limit to the amount synthesized becomes exponentially smaller for larger values of $d$ and the corresponding time required to synthesize a fixed amount of the molecular chain becomes exponentially longer \cite{Note2}.

\footnotetext[2]{See Supplemental Material for a detailed derivation.}
\footnotetext[3]{Strictly speaking, this inequality cannot be applied to all connections, which results in a factor of $1/3$ in Eq. (\ref{T_gousei}). Please refer to Supplemental Material for details.}

% 合成が線形時間で済むことの証明の流れ
Next, we outline the proof that the decomposition time is shorter than a linear function of the molecular chain length. Although it is possible to directly solve Eqs. (\ref{bunkai_rate_eq_M}) and (\ref{bunkai_rate_eq_X}), we instead  consider the probability that the molecular chain $M$ is completely degraded within time $t$. Using Markov's inequality, we find that $c_X(t)$ has a lower bound, which leads to Eq. (\ref{T_bunkai}) \cite{Note2}.

% 非対称性を生み出す機構について
\SECTION{Mechanism behind the hard synthesis and easy destruction}
The asymmetry exhibited by our model can be understood from its reaction network structure, which is a Cayley graph of depth $n$ (See Fig. \ref{fig:Cayley}). Because of byproduct formation, the proportion of molecules that can reach the desired vertex ($\EXAMPLE$ in this case) becomes exponentially smaller than the initial amount.  As a result, the synthesis rate is markedly reduced. 
In contrast, during chain decomposition, no byproducts are produced and the reaction proceeds along a single path. Thus, decomposition only requires a time that is linear to the chain length.

\FIG{Cayley_graph}{fig:Cayley}
{Reaction network in our model. (a) Schematic illustration of the chain synthesis. The color intensity represents the number of molecules. The number of molecules decreases exponentially with the chain length because of the inevitable formation of byproducts. (b) Schematic illustration of chain decomposition. In this case, there is no byproduct, and the decomposition proceeds along a single path (red solid line).}

Although we have focused on a specific example, the scheme shown in Fig. \ref{fig:Cayley} --- an exponentially branched maze with only one exit --- may capture a universal aspect that can be applied to more complex and general asymmetries of operational difficulty.

% いくつかのコメント　---------------------------------
\SECTION{Concluding remarks}
% 結果の要約
We have described a simple model of the synthesis and decomposition of molecular chains composed of two types of molecular subunits.
As shown in Eq. (\ref{T_gousei}), the difficulty of synthesizing heteropolymers in this system is characterized by the number of connections $d$ between different types of monomers.
Equations (\ref{T_gousei}) and (\ref{T_bunkai}) highlight the distinct asymmetry between assembly and destruction.
When $d$ is proportional to $n$, the time required to assemble a fixed amount of molecular chains is an exponential function of the chain length regardless of how the concentrations of the monomers are manipulated, while the corresponding function for chain destruction is linear. 
The result in this {\PAPER} is located at the starting point for the study of a general question about the hardness of operations. We thus expect that the exact form of the bound given in this {\PAPER} would be useful for seeking a general principle.

% 限界
In closing, we present five future challenges.
First, this study ignored the reverse reaction for simplicity, but we expect from the following naive discussion that explicitly considering the reverse reaction would not affect the results. In the present model, molecular chain $MAA$ is inevitably generated as a byproduct of $MAB$ creation and is the cause of the slow synthesis. If we included the reverse reaction (detachment of $A$), we would reduce the amount of byproducts but also reduce the amount of the desired product, $MAB$.  Therefore, we would not accelerate the rate of molecular synthesis. The analysis of the model with a reverse reaction is left as a future task. 

% 熱力学的不確定性関係・速度限界との関係性
A second problem is related to stochastic thermodynamics, where an upper bound of the current in nonequilibrium steady-state or a lower bound of the operation time to convert a probability distribution to another one is intensively studied as thermodynamic uncertain relation \cite{Barato_2015, Horowitz_2020} and thermodynamic speed limit \cite{Shiraishi_2018, Ito_2020}. Although the settings are different in these relations and the results presented in this {\PAPER}, they share the same motivation of searching for a limit that cannot be exceeded regardless of the operational protocol. Thus, it is natural for future research to explore the relationship among them by extending the present results to stochastic systems.

% エントロピーとの関係について
A third problem is related to entropy.
One may expect that our result is obtained from a fact that the thermodynamic entropy of a molecular chain is lower than that of disconnected monomer units. While it is impossible to make a lower entropy state from a higher one in thermally isolated systems, the system we study operates out of equilibrium. Thus, the thermodynamic entropy is not directly related to the asymmetry of operational hardness. 
Rather, our results suggest that it is not the entropy of the molecular chain but its blockiness that characterizes the efficiency of heteropolymer synthesis.
When $d$ is small, i.e., the heteropolymer contains large $A (B)$ clumps, the desired molecular chain can be synthesized relatively efficiently by controlling the concentrations of the monomers.
Conversely, when $d$ is large, i.e., the heteropolymer does not contain large clumps of $A (B)$, efficient synthesis cannot be achieved no matter how the concentrations of the monomers are varied over time.
Although our results are limited to one-dimensional heteropolymers, this observation may be useful to explore the fundamental limits on the synthesis of more general polymers or proteins under nonequilibrium environments.

% フィードバック制御および情報熱力学との関係
A fourth problem is related to feedback control.
Information thermodynamics \cite{Sagawa_2008, Parrondo_2015, Toyabe_2010}, which extends thermodynamics to include the effects of measurement and feedback control by using information-theoretic quantities, was formulated in the last decade.
The system discussed in this {\PAPER} is described by deterministic rate equations and therefore does not include the concept of feedback. However, in small fluctuating systems, error correction by feedback mechanisms is expected to be important to generate complex objects \cite{Juarez_2012, Rothemund_2004}.
Extending our results to stochastic systems where feedback effects become significant is an important research direction to discuss structure formation at small scales.

% 実験との関係
Finally, we remark that the synthesis of one-dimensional molecular chains has been experimentally studied \cite{Rubinstein_2003, Howker_2005, Mast_2013, Cohen_2013, Korevaar_2012}. Since the base of the exponential function in Eq. (\ref{T_gousei}), $2^{1/3}$, is a model-specific parameter, its experimental determination may yield information about the underlying chemical reactions. Formulating a general relationship between chemical reaction networks and operational hardness represents the most significant future challenge.  

% 謝辞
We thank Andreas Dechant, Masato Itami, and Tomohiro Tanogami for fruitful discussions.
This work was supported by KAKENHI (Grant Nos. 17H01148, 19H05795, and 20K20425).

% BibTeX で参考文献リストを自動生成する
\bibliographystyle{myaps}
\bibliography{bib}

% 導出の詳細　---------------------------------------------------------------------
\SM{Detailed derivation}

% 合成に指数時間を要することの証明
\subsection*{Derivation of Eq. (\ref{T_gousei})}
To derive Eq. (\ref{T_gousei}), we first consider the sequential addition of molecules $A$, $B$, and $Z^*$ to the chain $MZ$ to synthesize $MZABZ^*$  ($Z,Z^*\in \SET{A,B}$). \\
The rate equations are
\ARRAY{lllll}{ \label{gousei_ODE}
    \DIFF{t} c_{MZ}(t) &=& -w(t)c_{MZ}(t) &+& w_Z(t)c_M(t), \\[10pt]
    \DIFF{t} c_{MZA}(t) &=& -w(t)c_{MZA}(t) &+& w_A(t)c_{MZ}(t), \\[10pt]
    \DIFF{t} c_{MZAB}(t) &=& -w(t)c_{MZAB}(t) &+& w_B(t)c_{MZA}(t),\\[10pt]
}
where $w_Z(t) \equiv k_Zc_Z(t)$ and $w(t) \equiv w_A(t) + w_B(t)$.
Solving the differential equations $(\ref{gousei_ODE})$ under the initial condition $c_{MZ}(0) = c_{MZA}(0) = c_{MZAB}(0) = 0$, we obtain
\EQ{ \label{c_MZAB}
c_{MZAB}(t) = \INT{0}{t}{dt_3}{w_B(t_3)} \INT{0}{t_3}{dt_2}{w_A(t_2)} \INT{0}{t_2}{dt_1}{w_Z(t_1)c_M(t _1)}  \EXP{-\INT{t_1}{t}{d\tau}{w(\tau)}}.
}
From Eq. (\ref{Sigma_def}), we have
\EQ{ \label{Sigma_MZABZ*}
\synGAIN{MZABZ^*}{0}{T} = \INT{0}{T}{dt}{w_{Z^*}(t) c_{MZAB}(t)}.
}
We then substitute Eq. (\ref{c_MZAB}) into Eq. (\ref{Sigma_MZABZ*}) and transform the integration variables as
\ARRAY{l}{
    s_1 = t_1,\\
    s_2 = t_2-t_1,\\
    s_3 = t_3-t_1,\\
    s_4 = t-t_1,
}
where the domain of integration is $D_s = \{(s_1,s_2,s_3,s_4) \,|\, 0 \leq s_1\leq T, 0\leq s_2\leq  s_3 \leq s_4\leq T-s_1\}$.
Because the corresponding Jacobian is unity, we have
\EQ{ \label{long}
\synGAIN{MZABZ^*}{0}{T} =
\INT{0}{T}{ds_1}{w_Z(s_1)c_M(s_1)} \INT{0}{T-s_1}{\hspace{-15pt}ds_4}{w_{Z^*}(s_1+s_4)} e^{-W(s_4; s_1)} 
\underline{ % この部分だけを取り出して、さらに不等式評価を続ける
\INT{0}{s_4}{ds_3}{w_B(s_1+s_3)} \INT{0}{s_3}{ds_2}{w_A(s_1+s_2)}
},}
where $W(s_4; s_1)\,\equiv \int_0^{s_4} du\, {w(s_1+u)}$. Using $w_A(s_1+s_2)\geq0$ and $s_3\leq s_4$, we evaluate the underlined part of Eq. (\ref{long}) as
\EQ{ \label{AM_GM_1}
\INT{0}{s_4}{ds_3}{w_B(s_1+s_3)} \INT{0}{s_3}{ds_2}{w_A(s_1+s_2)}
\leq \INT{0}{s_4}{ds_3}{w_B(s_1+s_3)} \INT{0}{s_4}{ds_2}{w_A(s_1+s_2)}.
}
Here, by applying the inequality of arithmetic and geometric means, we find that the right-hand side is less than
\EQ{ \label{AM_GM_2}
\FRAC{1}{4} \qty(\INT{0}{s_4}{ds_3}{w_B(s_1+s_3)} + \INT{0}{s_4}{ds_2}{w_A(s_1+s_2)})^2, }
which is written as
\EQ{ \label{AM_GM_3}
\FRAC{1}{4} W(s_4; s_1)^2.
} % 下線部の不等式評価が完了

Substituting Eqs. (\ref{AM_GM_1}), (\ref{AM_GM_2}), and (\ref{AM_GM_3}) into Eq. (\ref{long}) and using $w \geq w_{Z^*}$, we obtain
\EQ{
\synGAIN{MZABZ^*}{0}{T} \leq
\FRAC{1}{4} \INT{0}{T}{ds_1}{w_Z(s_1)c_M(s_1)} \INT{0}{T-s_1}{\hspace{-15pt}ds_4}{w(s_1+s_4)} W(s_4; s_1)^2 e^{-W(s_4; s_1)} .}
From the definition of $W(s_4; s_1)$, $ds_4 w(s_1+s_4)$ is equal to $ dW(s_4; s_1)$. We then obtain 
\EQ{ \label{half_AB_1}
\synGAIN{MZABZ^*}{0}{T}
\leq \FRAC{1}{4}\INT{0}{T}{ds_1}{w_Z(s_1)c_M(s_1)} \INT{0}{W(T-s_1; s_1)}{\hspace{-10pt}dW}{W^2 e^{-W}}.
}
Because $W^2 e^{-W} \geq 0$ and $0 \leq W(T-s_1; s_1) < \infty$, the last part of Eq. (\ref{half_AB_1}) is evaluated as
\EQ{ \label{gamma_2}
\INT{0}{W(T-s_1; s_1)}{\hspace{-10pt}dW}{W^2 e^{-W}}
< \INT{0}{\infty}{dW}{W^2 e^{-W}} = 2.
}
By combining Eq. (\ref{half_AB_1}) and Eq. (\ref{gamma_2}), we obtain
\EQ{ \label{half_AB}
\synGAIN{MZABZ^*}{0}{T} < \FRAC{1}{2} \synGAIN{MZ}{0}{T}.
}

Because this model is symmetric with respect to the interchange of molecules $A$ and $B$, we also obtain
\EQ{ \label{half_BA}	
\synGAIN{MZBAZ^*}{0}{T} < \FRAC{1}{2}\synGAIN{MZ}{0}{T}.
}

\FIG{half}{fig:half}
{Schematic diagram of the proof of the synthesis process.
When $A, B$ and $Z^* (Z^* \in \SET{A, B})$ are sequentially connected to the molecular chain $MZ$, no matter how $w_A(t)$ and $w_B(t)$ are manipulated, $\synGAIN{MZABZ^*}{0}{T}$ cannot exceed the half of $\synGAIN{MZ}{0}{T}$.
Note that at most three connections between $A$ and $B$ are consumed in order to perform this evaluation once. Indeed, the number of connections in $MB$ is three less than that in $MBABA$ (corresponding to the case where $Z=B, Z^*=A$).
Therefore, if the number of connections between $A$ and $B$ in the target chain is $d$, then this inequality (or swapping version of $A$ and $B$) can be used at least $d/3$ times, and the maximum synthesis amount decreases exponentially by $2^{-d/3}$.}

Now, suppose that the molecular chain $M$ contains $d$ times of connections between $A$ and $B$, we attempt to apply Eq. (\ref{half_AB}) and Eq. (\ref{half_BA}) iteratively.
For example, in Fig. (\ref{fig:iteration}), we show how the inequalities work for a sequence.
As seen in the example, Eq. (\ref{half_AB}) or Eq. (\ref{half_BA}) can be used at least $d/3$ times.

\FIG[width=4.0cm]{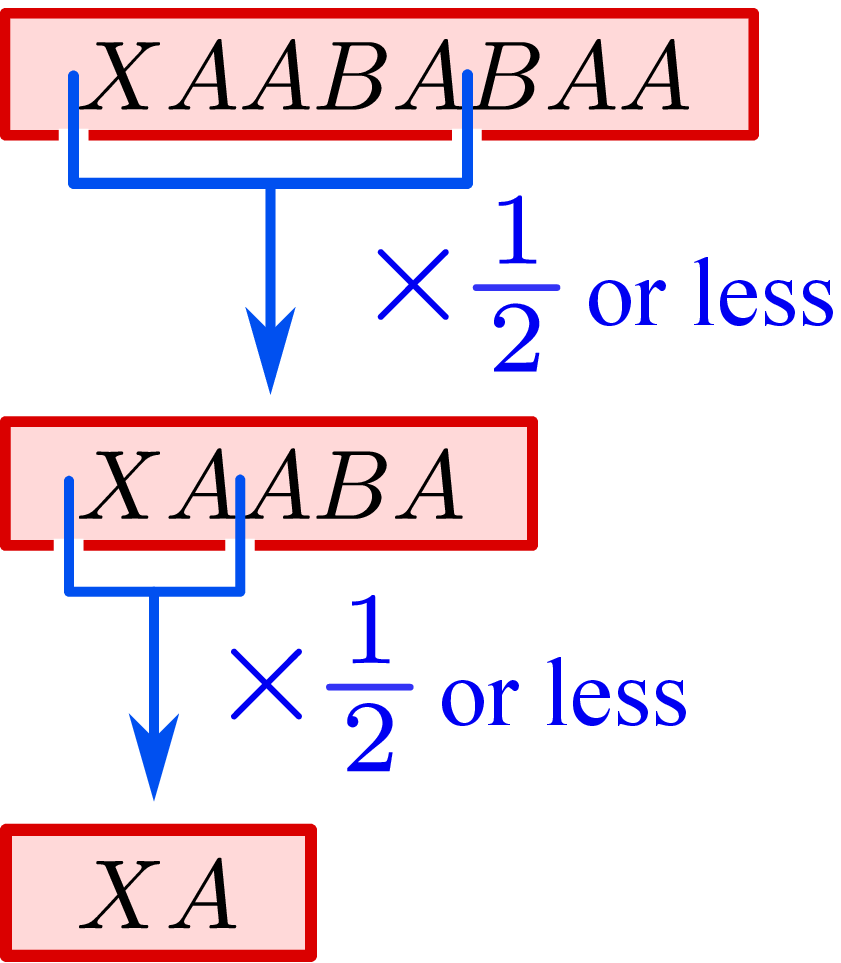}{fig:iteration}
{Example of iterative application of Eq. (\ref{half_AB}) or Eq. (\ref{half_BA}).
In this example, either Eq. (\ref{half_AB}) or Eq. (\ref{half_BA}) is iteratively applied to the molecular chain $XAABABAA$.
In the first step, by applying Eq. (\ref{half_BA}) with $MZ=XAAB$ and $Z^*=A$, we find that $\synGAIN{XAABABAA}{0}{T} < \synGAIN{XAABA}{0}{T}\times 1/2$.
Next, by applying Eq. (\ref{half_AB}) with $MZ=XA$ and $Z^*=A$, we find that $\synGAIN{XAABA}{0}{T} < \synGAIN{XA}{0}{T}\times 1/2$. 
In this example, $n=7, d=4$, and one of Eqs. (\ref{half_AB}) and (\ref{half_BA}) is applied twice, which is larger than $d/3 = 1.333...$ times.
}

We thus obtain
\EQ{ \label{exp_small}
 \synGAIN{M}{0}{T} < 2^{-d/3}\times \synGAIN{XZ_1}{0}{T},}
where $Z_1$ is the molecular subunit connected to the reaction nucleus $X$ ($Z_1 \in \SET{A,B}$).

Finally, we evaluate $\synGAIN{XZ_1}{0}{T}$.
Eq. (\ref{gousei_rate_eq_X}) is written as
\EQ{ \label{gousei_rate_eq_X_revised}
\DIFF{t}c_X(t) = -w(t)c_X(t) + J_X.
}
From Eq. (\ref{Sigma_def}), we have
\EQ{ \label{Sigma_XZ1} 
\synGAIN{XZ_1}{0}{T} = \INT{0}{T}{dt}{w_{Z_1}(t) c_{X}(t)}. 
}
By using $w_{Z_1} \leq w$, the right-hand side of Eq. (\ref{Sigma_XZ1}) is less than
\EQ{ \label{wc_X}
\INT{0}{T}{dt}{w(t) c_{X}(t)}.
}
By integrating Eq. (\ref{gousei_rate_eq_X_revised}) from $t=0$ to $t=T$, we obtain
\ARRAY{lll}{ \label{int_rate_eq_X}
\INT{0}{T}{dt}{w(t) c_{X}(t)} 
&=& \INT{0}{T}{dt}{\qty(J_X - \DIFF{t}c_{X}(t))}\\[10pt]
&=& J_XT - c_X(T) + c_X(0).}
Because $c_X(0) = 0$ and $c_X(T) \geq 0$, the last line is less than
\EQ{ \label{J_XT} J_XT.}
From Eqs. (\ref{Sigma_XZ1}), (\ref{wc_X}), (\ref{int_rate_eq_X}), and (\ref{J_XT}), we obtain
\EQ{ \label{eval_Sigma_XZ1}
\synGAIN{XZ_1}{0}{T} \leq J_XT.
}

Combining Eq. (\ref{exp_small}) and Eq. (\ref{eval_Sigma_XZ1}), we obtain
\EQ{ \label{Sigma_gousei_exp}
\synGAIN{M}{0}{T} < 2^{-d/3}\times J_XT,}
which is equivalent to Eq. (\ref{T_gousei}).

% 分解が線形時間で済むことの証明
\subsection*{Derivation of Eq. (\ref{T_bunkai})}
Next, we derive Eq. (\ref{T_bunkai}).
Let $p_n(t)$ be the probability distribution of the time it takes for a molecular chain $M = XZ_1Z_2\cdots Z_n$ of length $n$ to be completely degraded. Then
\ARRAY{lll}{
c_X(t)
&=& \INT{0}{t}{dt'}{c_M p_n(t-t')}\\[10pt]
&=& c_M \INT{0}{t}{dt'}p_n(t')\\[10pt]
&\equiv& c_M P_n(t),
}
where $P_n(t)$ is the cumulative distribution function of $p_n(t)$. Using Markov's inequality, we evaluate $P_n(t)$ as
\ARRAY{lll}{
P_n(t)
&>& 1-\FRAC{\AVE{t}_{p_n}}{t}\\[8pt]
&=& 1-\FRAC{1}{t} \SUM_{i=1}^{n} \FRAC{1}{k'_{Z_i}} \\[15pt]
&=& 1-\FRAC{\overline{\tau}n}{t},
}
where 
\EQ{\overline{\tau} \equiv \FRAC{1}{n} \SUM_{i=1}^{n} \FRAC{1}{k'_{Z_i}}}
is the average molecular detachment time.  

Because $c_X(t)$ is a monotonically increasing function of $t$, we can evaluate $ \int_0^T dt\,{c_{X}(t)}$ as shown in FIG. \ref{fig:cX(t)} to obtain
\ARRAY{lll}{ \label{eval_cX(t)}
\decGAIN{X}{0}{T}
&=& \INT{0}{T}{dt}{\tau_X^{-1} c_{X}(t)}\\[15pt]
&>& \tau_X^{-1} c_X(t')\qty(T-t')\\[10pt]
&>& \tau_X^{-1} c_M \qty(1-\FRAC{\overline{\tau}n}{t'})\qty(T-t').
}
for $0<t'<T$.
Setting $t' = 2\overline{\tau}n$, we obtain Eq. (\ref{T_bunkai}).

\FIG[width=6.0cm]{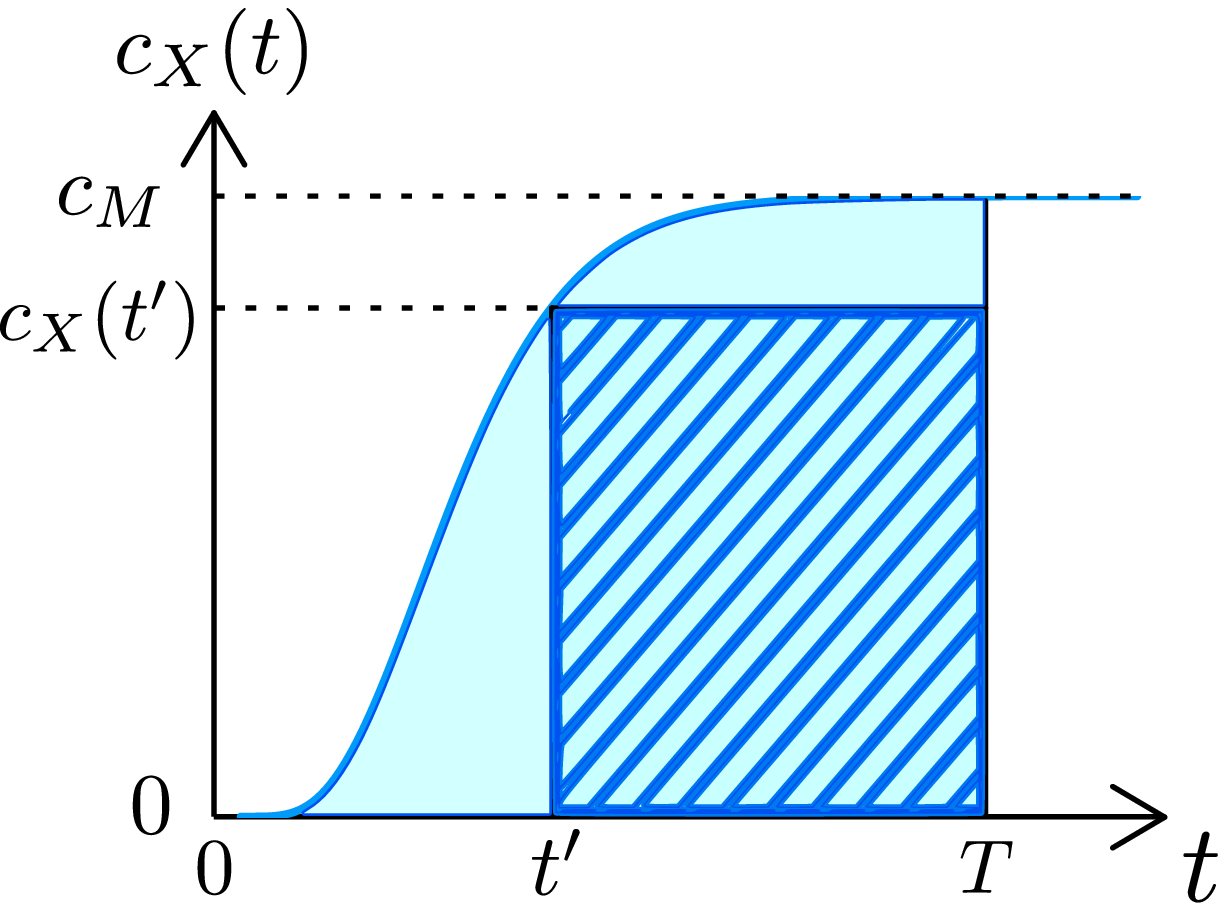}{fig:cX(t)}{
Schematic illustration of Eq. (\ref{eval_cX(t)}).
The area of painted in light blue $ \int_0^T dt\,{c_{X}(t)}$ is larger than the area of the shaded rectangle $c_X(t')\qty(T-t')$.
}

\end{document}